\documentclass[10pt, conference]{IEEEtran}
\usepackage[utf8]{inputenc}
\usepackage[hidelinks,bookmarks=false]{hyperref}
\usepackage{amssymb}
\usepackage{amsmath}
\usepackage{bm}
\usepackage{cancel}

\usepackage{graphicx}
\usepackage{xcolor}
\usepackage{import}
\usepackage{listings}
\usepackage[noend]{algpseudocode}
\usepackage{cite}

\usepackage[ruled,vlined]{algorithm2e}
\usepackage{paralist}
\usepackage{textcomp}
\usepackage{tikz}
\usepackage{lipsum}
\usepackage{tabularx}
\usepackage{pgfplots}
\usepackage[normalem]{ulem}
\usepackage{booktabs}

\usepackage{subcaption}

\pgfplotsset{compat=newest}
\usepgfplotslibrary{units}

\algrenewcommand\algorithmiccomment[1]{\hfill \textcolor{gray}{$\triangleright$ \textit{#1}}}

\renewcommand{\paragraph}[1]{\vspace{0.1cm}\noindent{\bf #1.}}

\newcommand{\myparagraph}[1]{\vspace{0.1cm}\noindent{\textbf{#1.}}}

\newcommand{\name}{PoS-CoPOR\xspace}

\begin{document}

\title{\name: Proof-of-Stake \underline{Co}nsensus \underline{P}rotocol with Native \underline{O}nion \underline{R}outing Providing Scalability and DoS-Resistance}

\author{
    \IEEEauthorblockN{
        Ivan Homoliak\IEEEauthorrefmark{1}\IEEEauthorrefmark{2},
        Martin Perešíni\IEEEauthorrefmark{1},
        Marek Tamaškovič\IEEEauthorrefmark{1},
        Timotej Ponek\IEEEauthorrefmark{1},
        Lukáš Hellebrandt\IEEEauthorrefmark{1} and
        Kamil Malinka\IEEEauthorrefmark{1}
    }
    \IEEEauthorblockA{
        \IEEEauthorrefmark{1}Brno University of Technology, Brno, Czech Republic \\
        \IEEEauthorrefmark{2}Slovak Technical University, Bratislava, Slovak Republic
        }
    }

\maketitle

\begin{abstract}
Proof-of-Stake (PoS) consensus protocols often face a trade-off between performance and security.
Protocols that pre-elect leaders for subsequent rounds are vulnerable to Denial-of-Service (DoS) attacks, which can disrupt the network and compromise liveness.
In this work, we present \name, a single-chain PoS consensus protocol that mitigates this vulnerability by integrating a native onion routing mechanism into the consensus protocol itself.
\name combines stake-weighted probabilistic leader election with an anonymization layer that conceals the network identity of the next block proposer.
This approach prevents targeted DoS attacks on leaders before they produce a block, thus enhancing network resilience.
We implemented and evaluated \name, demonstrating its ability to achieve a throughput of up to 110 tx/s with 6 nodes, even with the overhead of the anonymization layer.
The results show that native anonymization can provide robust DoS resistance with only a modest impact on performance, offering a solution to build secure and scalable PoS blockchains.
\end{abstract}

\section{Introduction}
\label{sec:intro}
In recent years, Proof-of-Stake (PoS) consensus protocols have started to proliferate due to their low utilization of energy resources as opposed to Proof-of-Work protocols. 
However, they might have several issues and limitations. 
In many of PoS protocols, it is possible to perform a DoS attack on the leader of the round since she is known beforehand. 
This might have two consequences:
(1) the adversary might increase her chances of being elected as the leader while increasing her relative profit and 
(2) it might cause a decrease in throughput, and thus compromise the reputation of the blockchain (as happened, e.g., in Solana~\cite{2022-solana-ddos}). 

The typical representatives of DoS-vulnerable protocols are based on the Byzantine Fault Tolerance (BFT) algorithm~\cite{castro1999practical}. 
Such protocols do not scale in the number of consensus nodes, although they might provide a high transactional throughput.
Examples in this category are Tendermint~\cite{tendermint,buchman2016tendermint}, Solana~\cite{2017-solana-whitepaper}, Harmony~\cite{2022-harmony-whitepaper}  with claimed throughputs of $10k$ tx/s, $50k$ tx/s, and $2k$ tx/s,  respectively. 
Another vulnerable example is Ethereum 2.0, which uses Randao-based leader election~\cite{EthereumPoSAttackDefense}.
Whisk~\cite{Whisk} was proposed as a defense technique; however, some variants of DoS still exist~\cite{2025-Tereza-DoS}.

On the other hand, Algorand~\cite{algorand} is a consensus protocol resistant to DoS on the leader.
Algorand uses a committee-based adaptation of BFT that enables scalability in the number of consensus nodes and, at the same time, provides fast transactional throughput (favoring consistency over availability).
Algorand is secure against the DoS on the leader because it uses a Verifiable Random Function (VRF)-based~\cite{vrf} leader/committee election algorithm that does not reveal the leader/committee before the round starts.
Algorand~\cite{algorand} has reported a maximum throughput of 3k txs/s.
However, independent tests have shown that Algorand's actual throughput is significantly lower -- ranging from 200 to 1000 txs/s~\cite{algorand-200}. 
Nevertheless, this holds only in the optimistic case, while in the case of disagreement within the elected committee, the block round has to be restarted (to preserve consistency), which might further slow down the throughput.

Another related work focusing on the design of a secure PoS protocol with high throughput and short time to finality is LAKSA~\cite{reijsbergen2020laksa}.
The authors of LAKSA focused on stochastic leader election, who is committed in the follow-up rounds by a stochastically elected committee.
The authors report a throughput of 450--1300 txs/s when assuming transactions of size 450B.
However, LAKSA is not resistant to the DoS on the leader, and therefore, the authors discuss the use of anonymization protocols at the network layer.

\myparagraph{Proposed Approach}
We propose \name, a scalable Proof-of-Stake (PoS) protocol designed to resist Denial-of-Service (DoS) attacks on the leader while offering moderate transactional throughput.
\name emulates Bitcoin's ability to scale consensus nodes by using a stochastic leader election that reflects diverse stakeholder power.
Unlike Bitcoin~\cite{nakamoto2008bitcoin}, where the leader is known post-block production, \name's leader is identified one round in advance.
To prevent DoS attacks, we integrate a native anonymization protocol directly into the consensus mechanism.
This approach leverages existing strong connections and minimal network latency among most consensus nodes, reducing desynchronization risks and maximizing profits, all while achieving moderate throughput with minimal anonymization overhead.

\paragraph{\textbf{Contributions}}
The contributions of this work are as follows:
\begin{enumerate}
    \item We design and implement \name, a PoS protocol that integrates a native onion routing layer for DoS protection on leaders, exploring Tor-like, Gossip, and Dandelion routing.
    \item We evaluated \name's performance, achieving moderate throughput (up to 110 tx/s in a 6-node network) with manageable anonymization overhead in a particular scenario.
    \item We analyze \name's security, detailing its DoS resilience, liveness, safety, and limitations.
\end{enumerate}
\section{The \name Protocol}
\label{sec:protocol}

In this section, we propose \name (\underline{Co}nsensus \underline{P}rotocol utilizing \underline{O}nion \underline{R}outing), a new PoS consensus protocol that is resistant against DoS on the leader and at the same time provides moderate transactional throughput. 
We position our work in the network and consensus layer of the security reference architecture~\cite{homoliak2020security}.

\subsection{High-Level Overview}
We propose a PoS consensus protocol with a single leader of the round, who unambiguously determines the leader of the next round before the next round begins.
To avoid DoS attacks on the leader of the next round, we embed the anonymization of all participating nodes into the consensus protocol itself (i.e., into its network layer). 

At the beginning of the protocol, we assume a genesis list of all initial participants with the stake distributed by an initial coin offering (ICO) or another similar approach.
The order of the nodes in the list determines their IDs. 
Each node establishes peering connections to $n$ randomly selected nodes using the anonymization protocol (see \autoref{sec:details-anon}).

The participants then use the RandHound protocol~\cite{syta2017scalable} (or any other particular protocol that provides good distributed deterministic randomness) to produce a random seed $rand$.
This seed unambiguously determines the leader of the first round from all participants.
We ensure unambiguous leader selection via a combination of VRF-based randomness and stake-weighted probabilistic election.
This process iteratively generates a unique sequence of block leaders, known one round in advance, and protected from DoS attacks by network layer anonymization.

\subsection{Anonymization Layer}
\label{sec:details-anon}
We present different anonymization techniques and the process of joining the network, sending, and relaying messages (see also \autoref{anon-alg}).

\subsubsection{\textbf{Anonymization Types}}
Our initial idea was to implement an anonymization layer purely based on the TOR network concepts~\cite{tor}.
However, TOR focuses on client-server anonymization, whereas we require node-to-node anonymization.
This raised the question of whether to anonymize only traffic between nodeA and nodeB, or also outgoing traffic from nodeB.
The former risks easier deanonymization\footnote{Especially if anonymization circuits are not periodically changed.}, while the latter severely affects throughput.
To address this, we draw inspiration from the Dandelion approach for Bitcoin transaction gossiping~\cite{Dandelion,Dandelion++, dandelion_grin}, which uses pseudo-random gossiping to hinder traceability.

We consider the following anonymization techniques:

\begin{itemize}
    \item \textbf{TOR-like} - Each time the round leader creates the new block, she gossips it to all her peers via the pre-established anonymization circuits (the circuit to use is chosen randomly). 
    The last node in the circuit sends the block unencrypted to the addressee, making it look like the last node was the round leader for the adversary.
    The addressee then gossips the block again to her peers via the pre-established circuits.
    The need for onion encryption each time the message is gossiped can cause a decrease in throughput. 
    To avoid this issue, we propose two alternative anonymization types.

    \item \textbf{Gossip-node} - Here, each time the round leader creates the new block, she gossips it to all her last nodes in the pre-established anonymization circuits anonymously (via these circuits). 
    Here, the last node in the circuit is the addressee, so no unencrypted message is sent.
    The last node then gossips the received block encrypted by a pre-established symmetric key to all her peers.
    The peering nodes after the reception of the block gossip it in the same manner.
    Notice that here we removed the need for onion encryption for the subsequent gossiping of the block.
    Instead of onion encryption, we encrypt the message only with the symmetric key, to make it impossible for a traffic listening adversary to even know about the content of the message. 
    It is clear that the adversary-controlled node will be able to decrypt the message intended for her, but that is not of concern.

    \item \textbf{Dandelion} - The ``first round`` of block gossiping remains the same as in the \textbf{gossip-node} anonymization.
    However, compared to the previous anonymization type, the last node in the circuit sends the block unencrypted to her peering nodes.
    This removes additional encryption overhead, while still providing a reasonable level of anonymization (especially if the established circuits are changed for new ones periodically).
\end{itemize}

To make it even harder for the network traffic eavesdropping adversary to detect patterns in block gossiping, we could employ a mechanism to randomly decide when to stop gossiping messages via anonymization circuits and start sending them encrypted/unencrypted.\footnote{However, we have not implemented such a mechanism in the current paper.}
Further, as we have already mentioned, it is necessary to periodically replace established onion circuits with new ones to prevent the deanonymization of nodes caused by the usage of predictable routes for block gossiping.

\subsubsection{\textbf{Joining the Network}}
To join a network, a new node $A$ does the following:
\begin{enumerate}
    \item A new node $A$ receives a list of IP addresses of all nodes in a directory (its trustworthiness can be ensured in multiple ways, for example~\cite{directory}).

    \item $A$ selects $n$ sets of $m$ nodes and for each set builds a circuit consisting of $m-1$ selected nodes in sequential order $c_1, c_2, \ldots, c_{m-1}$ where $c_1$ is closest to $A$ in the circuit and $c_{m}$ is the consensus protocol peer. 
    To build a circuit, perform a key exchange with each of the selected nodes, for example, using the approach proposed in TOR~\cite{tor}.
	
    \item At this moment, $n$ circuits have been established with $n$ consensus protocol peers. 
    In each circuit, $A$ shares secret keys $K_i$ (under $\Sigma_{or}$) with consensus nodes $c_i$, where $i \in \{1, \ldots, m-1\}$. 
\end{enumerate}

\subsubsection{\textbf{Sending Messages}}
Any message $M$ that $A$ sends in the onion routing manner by each of $n$ circuits requires the following: 
\begin{enumerate}
    \item The message $M$ is encoded. 
    \item The message $M$ is encrypted with $K_m(M)$.
    \item The result of the previous step is encrypted with $K_x(.)$ for $x=m-1 \ldots 2$ and appended with the IP of the $(x)$th peer in the circuit. For example (m~=~4): $c_1, K_1(c_2,K_2(c_3,K_3(M)))$.
\end{enumerate}

\subsubsection{\textbf{Relaying Messages}}
\begin{itemize}
    \item When a peer $p_n$ in a circuit receives a message, it decrypts it using the key shared with $P$.
    Discovers the identity of $p_{n+1}$ and sends the message (which is still encrypted by $P$ using $K_{n+1}$) to it.
    \item If there is no $p(n+1)$, the peer is an exit peer.
    Decrypt the message and gossip about~\cite{jenkins2001gossip} it.
\end{itemize}

\begin{algorithm}[!t]
	\scriptsize
	\SetKwProg{func}{function}{}{}
	$\triangleright$ \textsc{Declaration of types and variables:}\\
	\hspace{1em} ${node:}$ \{ $addr$, $key$ \}, \\
	\hspace{1em} ${addr:}$ \{ $IP$, $port$ \}, \\
	\hspace{1em} ${this:}$ the current node, \\
	\hspace{1em} $routes:$ a list of routes that are used in anonymization layer, \\
	\hspace{1em} $Message:$ a constructor of a message, \\
	\smallskip
	
	\func{$joinNetwork$($n\_routes$, $m\_nodes$)} {
		$allnodes \leftarrow getNodes()$;\\
		$routes \leftarrow pickRoutes(n\_routes, m\_nodes)$;\\
		\For{$route$: $routes$}{
			\For{$node$: $route$}{
				$exchangeKey(node, this.key)$;\\
			}
		}
		\For{$route$: $routes$} {
			$buildCircuit(route)$;\\
		}
	}					
	\smallskip
	
	\func{$SendMessage$($dst$, $msg$)} {
		\For{$route$ : $routes$} {
			\For{$node$ : $route$} {
				$ct \leftarrow \Sigma_{or}.Encrypt(msg, node.key)$;\\
				$em \leftarrow Message("Encrypted", this.addr, ct)$;\\
				$relay\_msg \leftarrow Message("Relay", node.addr, em)$;\\
			}
			
			$ct \leftarrow \Sigma_{or}.Encrypt(relay\_msg, transport\_key)$;\\
			$send(msg.dst, Message("Transport", route[-1], ct))$;\\			
		}	
	}					
	\smallskip
	
	\func{$RelayMessage$($src$, $relay\_msg$)} {
		$msg\_key \leftarrow findKey(nodes, src)$;\\
		$msg \leftarrow \Sigma_{or}.Decrypt(relay\_msg, msg\_key)$;\\
		$transport\_key \leftarrow findKey(nodes, msg.dst)$;\\
		$ct \leftarrow \Sigma_{or}.Encrypt(msg, transport\_key)$;\\
		$send(msg.dst, Message(this.addr, ct))$;\\
	}					
	\smallskip
	
	\caption{Anonymization layer}\label{anon-alg}
\end{algorithm}

\subsection{Details of the Consensus Protocol}
To get the randomness for each round, we propose using VRF over the randomness from the previous round.
In detail, the randomness of the next round is obtained by the leader of the current round, who signs the randomness of the previous round.
This randomness can then be simply used to verify the leader it deterministically selects. 
Note that in contrast to Algorand, we do not use VRF in combination with a threshold for private selection of candidate leaders. 
Algorand's setting results in a selection of $0 \ldots n$ candidate leaders, which requires an additional voting round for the winner.

\paragraph{\textbf{Block Structure}}
Each block contains a header, the list of aggregated transactions, and the signature of the header made by the round leader.
The header of the block consists of the following fields:
\begin{itemize}
	\item \textbf{ID}: the counter of all blocks,
	\item \textbf{hPrev}: the hash of the previous block's header,
	\item \textbf{txsRoot}: the root of all transactions included in the block,
	\item \textbf{mptRoot}: the root hash of the Merkle Patricia trie aggregating all account states,
	\item \textbf{coinbase}: the PK of the node that is the leader of the block. It is used for signature verification,
	\item \textbf{rand}: the randomness of the current round, which determines the leader of the next round (and alternative leaders if the main one is not available). 
	This value represents a signature made by the leader of the current round on the randomness from the previous round (i.e., block),
	\item \textbf{altIdx}: is the index of the alternative leader who created the current block in the case a higher responsible leader was not available (see details later). 
	Hence, $altIdx = 0$ for the main leader of the round, $altIdx = 1, 2, 3, \ldots$ for the first alternative leader of the round, second one, etc.
	\item $\mathbf{\sigma}$: the signature of all the above fields in the header made by the creator of the block (i.e., the main leader or the first available alternative leader). 	
\end{itemize}
Each transaction of the block consists of the following fields:
\begin{itemize}
	\item \textbf{src}: the address of the sender of the transaction,
	\item \textbf{dst}: the address of the recipient of the transaction,
	\item \textbf{val}: the value sent from the sender to the recipient,
	\item \textbf{fee}: the fee that is payed to the leader who creates the block,
	\item \textbf{$\sigma$}: the signature of the transaction made by the sender.
\end{itemize}

\smallskip\noindent
Note that although our protocol can be used for smart contract platforms, in this work, we abstract from this use case, and for simplicity, we consider only the cryptocurrency use case (i.e., account states contain only balances and nonces, where the purpose of nonces is to avoid double-spending).

\paragraph{Initialization}
We assume that the code for the full node contains a list of the initial list of nodes whose stake is distributed (e.g., by ICO).
Although our protocol separates the balance from the stake (see \autoref{sec:incentives}), we assume that the initial distribution of all crypto tokens is put into the stake since each participant wants to participate in the consensus protocol.
Each node creates an anonymized connection to $N$ randomly selected nodes from the list of all nodes (see \autoref{sec:details-anon}), referred to as transport peers.
These peering nodes serve as transport relays for all messages and transactions that are sent/forwarded by the node.

Although our protocol can be used for smart contract platforms, in this work, we abstract from this use case, and for simplicity, we consider only the cryptocurrency use case (i.e., account states contain only balances and nonces, where the purpose of nonces is to avoid double-spending).

\begin{algorithm}[!t]
	\scriptsize
	\SetKwProg{func}{function}{}{}
	
	$\triangleright$ \textsc{Declaration of types and variables:}\\
	\hspace{1em} \textbf{Header} \{ $ID$, $hPrev$, $txsRoot$, $coinbase$, $rand$, $altIdx$, $\sigma$\}, \\
	\hspace{1em} \textbf{Block} \{ $hdr$, $txs$\}, \\
	\hspace{1em} \textbf{Tx} \{$src$, $dst$, $val$, $nonce$, $fee$, $\sigma$ \}, \\
	
	\smallskip
	\hspace{1em} this: the current node, \\
	\hspace{1em} gs:  the global state of the node contains \{$ID$, $PK$, $balance$, $nonce$, $stake$\} for each account state, \\
	
	\hspace{1em} $blocks$: mapping of block hashes to blocks,\\						
	\hspace{1em} $nodes$: all nodes with their IDs, PKs, stakes, and balances\\
	\hspace{1em} $mempool$: unprocessed txs,\\
	\hspace{1em} $BLast$: the hash of the last valid block,\\
	\hspace{1em} $R$: the number of expired timeouts of the round,\\	
	\smallskip
	
	\smallskip
	$\triangleright$ \textsc{Declaration of functions:}
	
	\func{$UponRoundStart()$} {				
		
		$R \leftarrow $ 0; \\		
		CreateBlock($R$); \\			
	}
	\smallskip	
	
	\func{$UponExprBlkTimeout$()} {
		$R \leftarrow R + 1$; \\ 		
		$CreateBlock(R)$; \\		
	}					
	\smallskip

	\func{$CreateBlock$($altIdx$)} {				
		
		\Comment{Check whether this node is the leader.\hfill \hfill} \\
		$leader, altLdrs \leftarrow$ Elect($blocks[BLast].hdr.rand$, $altIdx$); \\
		\If{$leader$ = nodes[this.ID]}{
			
			$txs \leftarrow pickBlkTxs(mempool)$; \\			
			
			$validateAndExecute(txs, gs)$; \Comment{Modify global state.} \\ 
			$Reward(leader, altLdrs, blk)$; \Comment{Rewarding of leaders.} \\	
			
			$newRand \leftarrow \Sigma_{pb}.Sign(this.SK, blocks.last().rand)$; \\
			$header \leftarrow (hPrev=BLast, 
			coinbase=this.PrK, 
			stRoot,
			rand=newRand, 
			txsRoot=txs, 
			ID=Block.ID+1, 
			altIdx=R)$; \\
			$\sigma \leftarrow \Sigma_{pb}.Sign(this.SK, h(header))$; \\
			$header.\sigma \leftarrow \sigma$; \\
			
			$blk \leftarrow (header, txs)$; \\
			$blocks$.add($blk$); \\
			gossip(blk);\\
			cancelTimeout(); \\ 
			
		}\Else{
			setBlkTimeout($\tau^B$); \\	
		}	
		
	}					
	\smallskip	
	
	\func{$UponRecvBlk$($blk$)} {
		\textbf{assert} $blk.hdr.hPrev = h(blocks[\text{BLast}].hdr)$; \\
		$\Sigma_{pb}.Verify((blk.hdr.coinbase, blk.hdr.\sigma), h(blk.hdr))$; \\
		
		\smallskip
		$leader, altLdrs \leftarrow Elect$($blocks[\text{BLast}]$.hdr.rand, $blk.hdr.altIdx$); \\
		\textbf{assert} blk.hdr.altIdx = $R$; \Comment{Check expired timeouts.} \\

		\textbf{assert} $leader.PK = blk.hdr.coinbase$; \Comment{Correct leader.} \\ 
		
		$\Sigma_{pb}.Verify((leader.PK, blk.hdr.rand), blocks[\text{BLast}].hdr.rand)$; \\ 
		
		$validateAndExecute(blk.txs, gs)$; \Comment{Modify global state.} \\
		\textbf{assert} $gs.stRoot = blk.hdr.stRoot$;\\
		$Reward(leader, altLdrs, blk)$; \\	
		
		$blocks$.append($blk$); \\
		cancelTimeout();
		
	}					
	\smallskip		
	
	\func{$UponRecvTx$($tx$)} {				
		\textbf{assert} $src \in nodes$; \\
		$\Sigma_{pb}.Verify((tx.src, tx.\sigma), h(tx.\{src, dst, val, nonce, fee\}))$; \\
		\textbf{assert} gs[tx.src].balance $\ge$ $tx.val$ + $tx.fee$; \\
		\textbf{assert} gs[tx.src].nonce = $tx.nonce$  ;\\
		$mempool.append(tx)$; \\
		gossip($tx$); \\
		
	}					
	\smallskip	
	
	\func{$Reward$($leader, altLdrs, blk$)} {
		
		\Comment{Reward the leader and $A$ alternative leaders. \hfill \hfill} \\
		$gs[leader.ID].balance ~\text{+=}~ Reward^F$; \Comment{Block reward} \\
		$gs[leader.ID].balance ~\text{+=}~ getTxFees(blk)$; \Comment{TX fees} \\
		
		\For{$n: altLdrs$}{
			$gs[n.ID].balance ~\text{+=}~ Reward^P$; \Comment{Partial reward} \\
		}	
		
	}					
	\smallskip
	
	\caption{Proposed consensus protocol.}
	\label{alg:consensus-protocol}
	\vspace{-0.1cm}
\end{algorithm}

\begin{algorithm}[t]
	\scriptsize
	\SetKwProg{func}{function}{}{}
	
	$\triangleright$ \textsc{Declaration of types and variables:}\\
	\hspace{1em} $MAX$: the size of interval used for election (e.g. 32B number),\\
	\hspace{1em} $ALT$: the desired number of alternative leaders\\
	\hspace{1em} $rand$: randomness for the current round (e.g. 32B number),\\
	\hspace{1em} $stakesum$: sum of all stakes in blockchain,\\
	
	\smallskip
	\func{$Elect$($rand$, $altIdx$)} {
		$slider \leftarrow 0$; \\
		\For{$node$: $nodes.sortByIDs()$}{
			$intervalSize \leftarrow \frac{node.stake~*~MAX}{stakesum}$; \\
			\smallskip
			$slider \leftarrow slider+intervalSize$; \\
			$intervalEnds[node.ID] \leftarrow slider$; \\
		}
		
		$leaders  \leftarrow []$; \\
		$leftToPick \leftarrow ALT + 1$; \\
		$currentRand \leftarrow rand$; \\
		\While{$leftToPick > 0$}{
			\For{$node,intervalEnd$: $intervalEnds$}{
				\If{$currentRand \le intervalEnd$}{
					$selectedNode \leftarrow node$; \\
					break; \\
				}
			}
			\If{$selectedNode \not \in leaders$}{
				$leaders.append(selectedNode)$; \\
				$leftToPick\text{-}\text{-}$; \\
			}
			$currentRand \leftarrow h(currentRand) \mod MAX$; \\
		}
		$leaders \leftarrow leaders[altIdx:]$; \\
		\Return{$(leaders[0],leaders[1:])$}; \\
		
	}					
	
	\smallskip
	\caption{Election function for leaders.}
	\label{alg:elect}
	\vspace{-0.1cm}
\end{algorithm}

\paragraph{Normal 
Operation}\label{sec:details-normal-operation}
The normal operation of the protocol is described in \autoref{alg:consensus-protocol}.
When the round starts, a node resets the counter $R$ of round-timeout expirations.
Then, a node checks whether it is the leader of the round based on the randomness from the previous round (see function $CreateBlock(0)$), and in the positive case it creates and gossips the block using a subset of transactions collected in the mempool.
In the negative case (i.e., it is not a leader), the node sets the synchronous timeout $\tau^B$ on the expiration of valid block delivery and it waits on the reception of the block (see function $UponRecvBlk()$). 
When a node receives a block, it makes a few validations: the binding of the block on its predecessor (i.e., $hPrev$), the signature of the header using the $coinbase$ field as the PK. 
Then, the correctness of the elected leader is verified by comparing the output of the function $Elect()$ with the $coinbase$ field.
In the positive case the rewarding of the leader and all alternative leaders takes place (see more details in \autoref{sec:incentives}).
Finally, the block is added into the ledger.

If a node does not receive a block within the timeout $\tau^B$, it indicates that the main leader is not available, and therefore the first available alternative leader must be selected and the round restarted (see details in \autoref{sec:detail-churn}).

If a node receives a transaction $tx$, it verifies the existence of the sender, her signature, enough balance, and correctness of the nonce.
Upon success, it adds $tx$ to the mempool of unprocessed transactions and gossips $tx$ to her peers.

\paragraph{Incentives and Rewarding Scheme}\label{sec:incentives}
We depict incentive scheme of our protocol in the function $Reward()$ of \autoref{alg:consensus-protocol}.
In detail, this function rewards the leader with the full reward $R^F$ and transaction fees, while $A$ alternatives leaders are rewarded with partial reward $R^P$.
This step decreases the variance of the reward, which is the problem of consensus protocols that reward only the leader of the round, such as in Bitcoin~\cite{blockchain}.
We emphasize that no already-rewarded participant can re-occur in the list of alternative leaders, and thus no double rewarding can occur within the same round.

In addition to the above, the rewarding of alternative leaders contributes to the liveness of the protocol -- the alternative leaders are incentivized in spreading the blocks across the network, even though they are not the leaders of those blocks.

\paragraph{\textbf{Stake and Balance}}
The important aspect of our rewarding scheme is that it separates the \textit{stake} from the \textit{balance} of the node, while the overall value of the node's assets is computed as their sum.
We define different rules for the stake and for the balance, each having its respective pros and cons.
The balance of the node has a high liquidity and thus can be used by the node to make any transactions according to her will.
Contrary, the stake of the node is meant to be a long-term investment that yields an ``interest rate``.
We emulate the long term investment and its interest rate by two requirements: 
\begin{compactitem}
	\item When a node wants to shift the part of its balance to its stake, the node has to wait $K^\#$ blocks before the stake is taken into account at function $Elect()$, thus enabling the node to earn profits. 
	This stake submission is made within a dedicated transaction.
	
	\item Similarly, when a node wants to shift a value from her stake to her balance by a transaction, then after the transaction is included, the node will have locked stake for $S^\#$ blocks before this is reflected, where $S^\#$ is big enough to penalize the liquidity of the node's crypto-assets.
	During the stake-locking period, the node does not obtain the profits from staking, which is to incentivize the staking time and mitigate a loss of cryptocurrency's market capitalization at the time of corrections.
\end{compactitem}

\begin{figure}[b]
    \vspace{-0.2cm}
    \centering
    \input{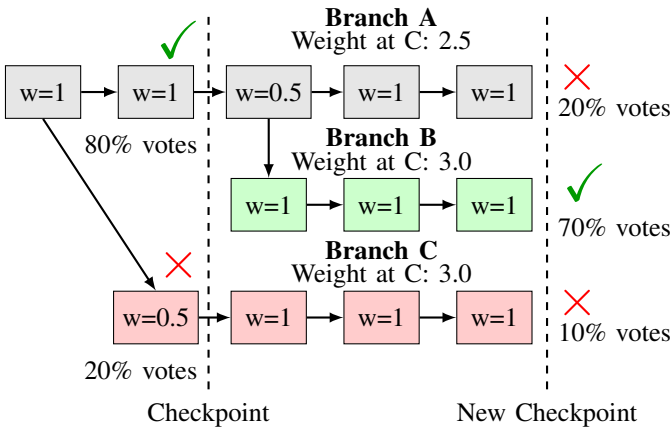}
    \vspace{-0.4cm}
    \caption{Check-pointing mechanism for resolution of forks. Two branches are depicted, with Branch B being selected due to higher cumulative weight and super-majority vote.}
    \label{fig:forks}
\end{figure}

\paragraph{Joining the Protocol}
To join the protocol, the node has to obtain/buy the balance from any of the existing nodes and then convert it into the stake.
Therefore, this permission model can be viewed as semi-permissionless~\cite{homoliak2020security}.

\subsubsection{\textbf{Churn of the Nodes}}\label{sec:detail-churn}
A node might temporarily go offline (i.e., as part of the churn phenomenon) and thus not produce a block when elected as the leader. 
If a node does not receive a block within the timeout $\tau^B$, then it increments the number of timeout restarts to $R = 1$ and checks whether the current node is not the first alternative leader entitled to create a block by calling the function $CreateBlock(R)$.
In the positive case, the node creates and gossips the block with adjusted $altIdx$ field to $R=1$.
In the negative case it resets the timeout $\tau^B$ and waits until a valid block is received from the next alternative leader determined by $R=2$.
If a valid block is not received within a timeout $\tau^B$, this procedure repeats again with incremented~$R$.

\paragraph{Context-Sensitive Transactions}
To disincentivize forking and thus improve safety, we employ the concept of context-sensitive transactions~\cite{gazi2018stake} that explicitly vote on a main chain they believe is the correct one by adding the hash of the recent block into the transaction itself. 
Therefore, they can be included (and their fees credited) only in such blocks that extend the voted chain.
More importantly, we propose to embed the votes of context-sensitive transactions into the chain quality computation, which will be described in \autoref{sec:forks}.
However, this mechanism adds an extra 32B size to the transaction, which is not crucial for the throughput of our protocol (see \autoref{sec:eval}).

\paragraph{Forks}\label{sec:forks}
The function $ChainQuality()$ serves to calculate the chain quality in the case of multiple chains occurring in parallel, forming inconsistencies called forks.
The highest contribution to the chain quality in a given chain is due to blocks created by the main leaders of rounds, while blocks created by alternative leaders contribute with decayed quality, depending on the index of the alternative leader (the higher the index, the lower the contribution).

Although there exists only a single ``highest-quality'' chain, which is created by the main leaders only, sometimes a main leader of the round might not be available (see \autoref{sec:detail-churn}), thus a block of the round is created by an alternative leader.
However, a node that was off-line during its ``leadership'' might return on-line and try to retrospectively create a block that could overturn the current strongest chain if it would be extended by the sequence of strong enough leaders.
This situation would negatively affect the finality and, more importantly, safety of the protocol.
To mitigate this situation, we add a bound on the finality by check-pointing after each $\mathcal{C}$ blocks, after which, all honest nodes do not accept overturning of the strongest chain.
Check-pointing is realized by 2/3 agreement in committee-based voting, while the committee is elected by VRF in the same manner as leaders.

The check-pointing and forks are illustrated in \autoref{fig:forks} with $\mathcal{C} = 3$, where $w$ represents a quality of each block computed within function $ChainQuality()$, which is called at the time of the checkpoints by all members of the committee.
In practice, the value of $C$ should correspond to a few minutes.
Periodical checkpoint also avoids stake bleeding attacks~\cite{gazi2018stake}.

\begin{figure*}[t]
	\centering
	\begin{subfigure}{0.32\textwidth}
		\includegraphics[width=\textwidth]{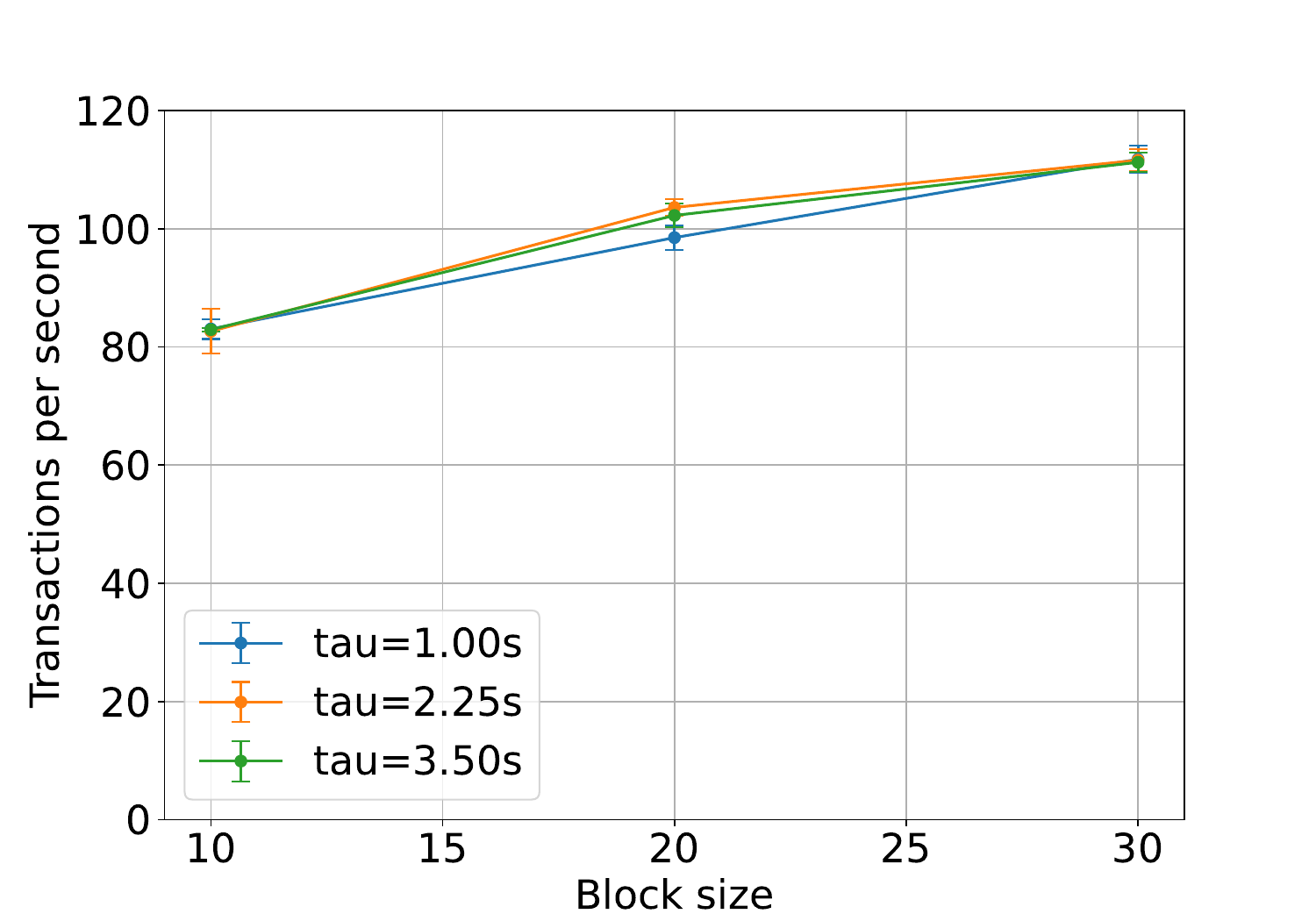}
		\caption{6/6 nodes - no anonymization}
	\end{subfigure}
	\begin{subfigure}{0.32\textwidth}
		\includegraphics[width=\textwidth]{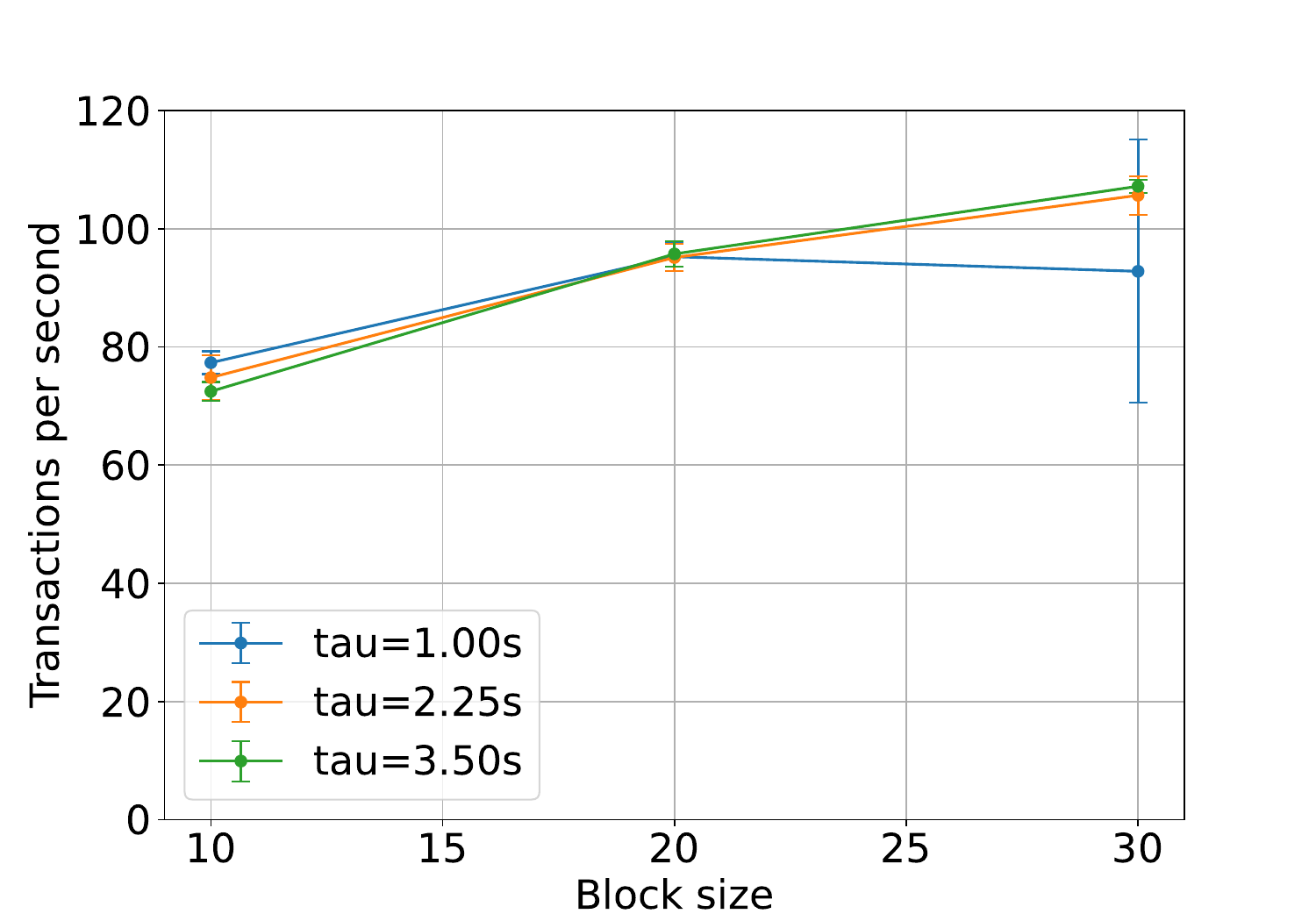}
		\caption{6/6 nodes - TOR anonymization}
	\end{subfigure}
	\begin{subfigure}{0.32\textwidth}
		\includegraphics[width=\textwidth]{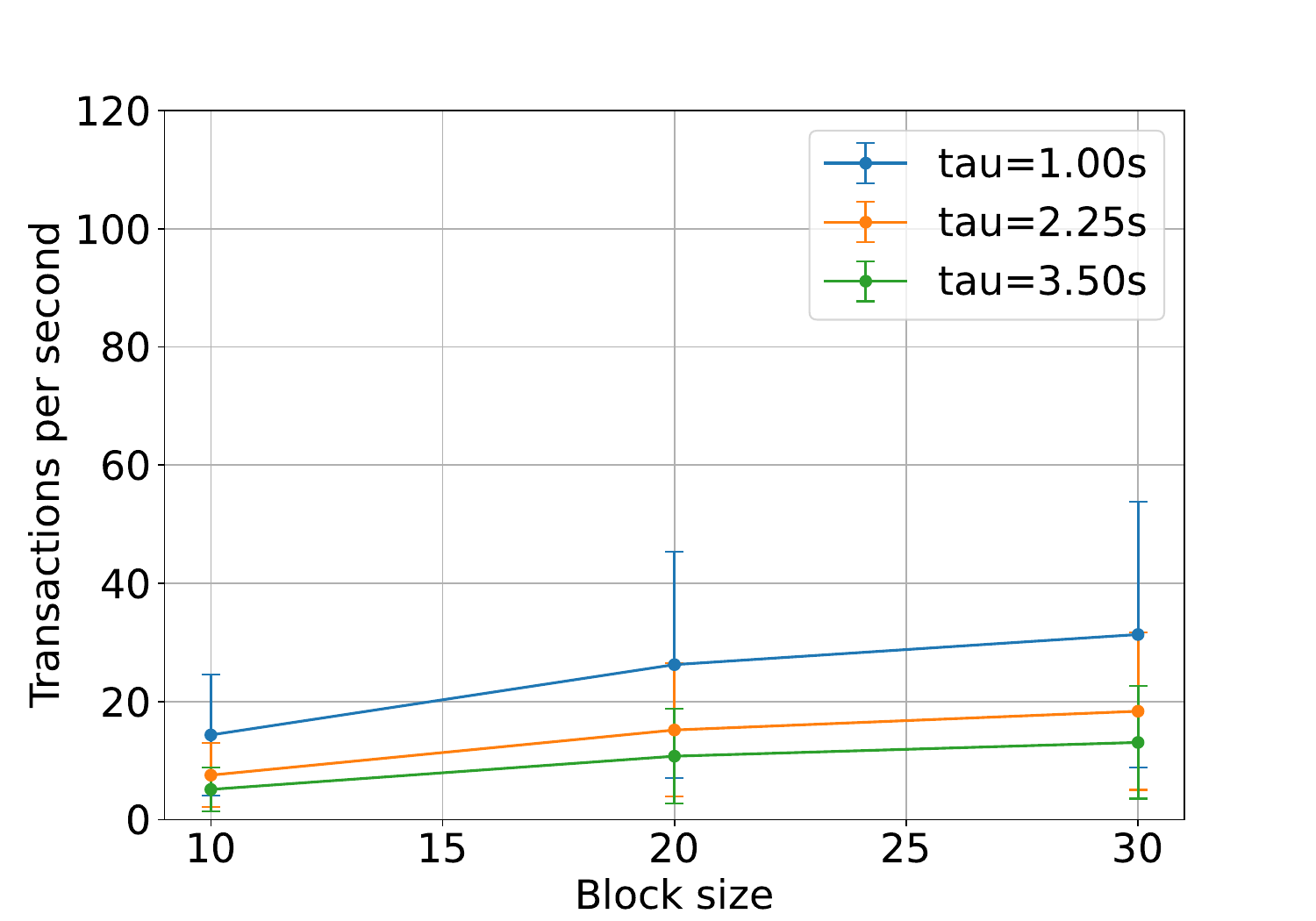}
		\caption{4/6 nodes - TOR anonymization}
	\end{subfigure}
	\caption{Selected experimental results with 6 nodes.}
	\label{fig:exp_nodes_6}
\end{figure*}

\begin{figure*}[!ht]
	\vspace{-0.4cm}
    \centering
	\begin{subfigure}{0.32\textwidth}
		\includegraphics[width=\textwidth]{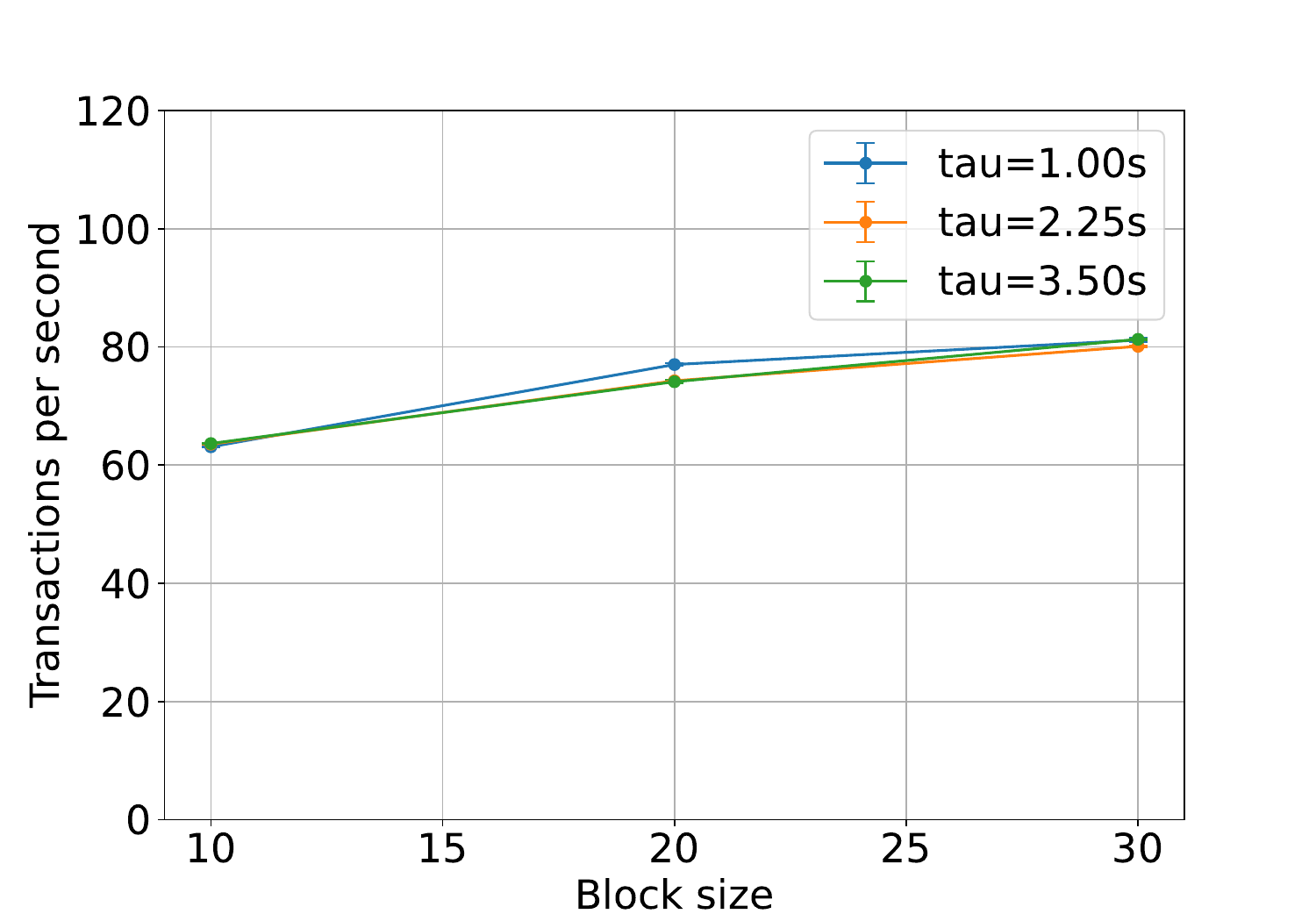}
		\caption{9/9 nodes - no anonymization}
	\end{subfigure}
	\begin{subfigure}{0.32\textwidth}
        \includegraphics[width=\textwidth]{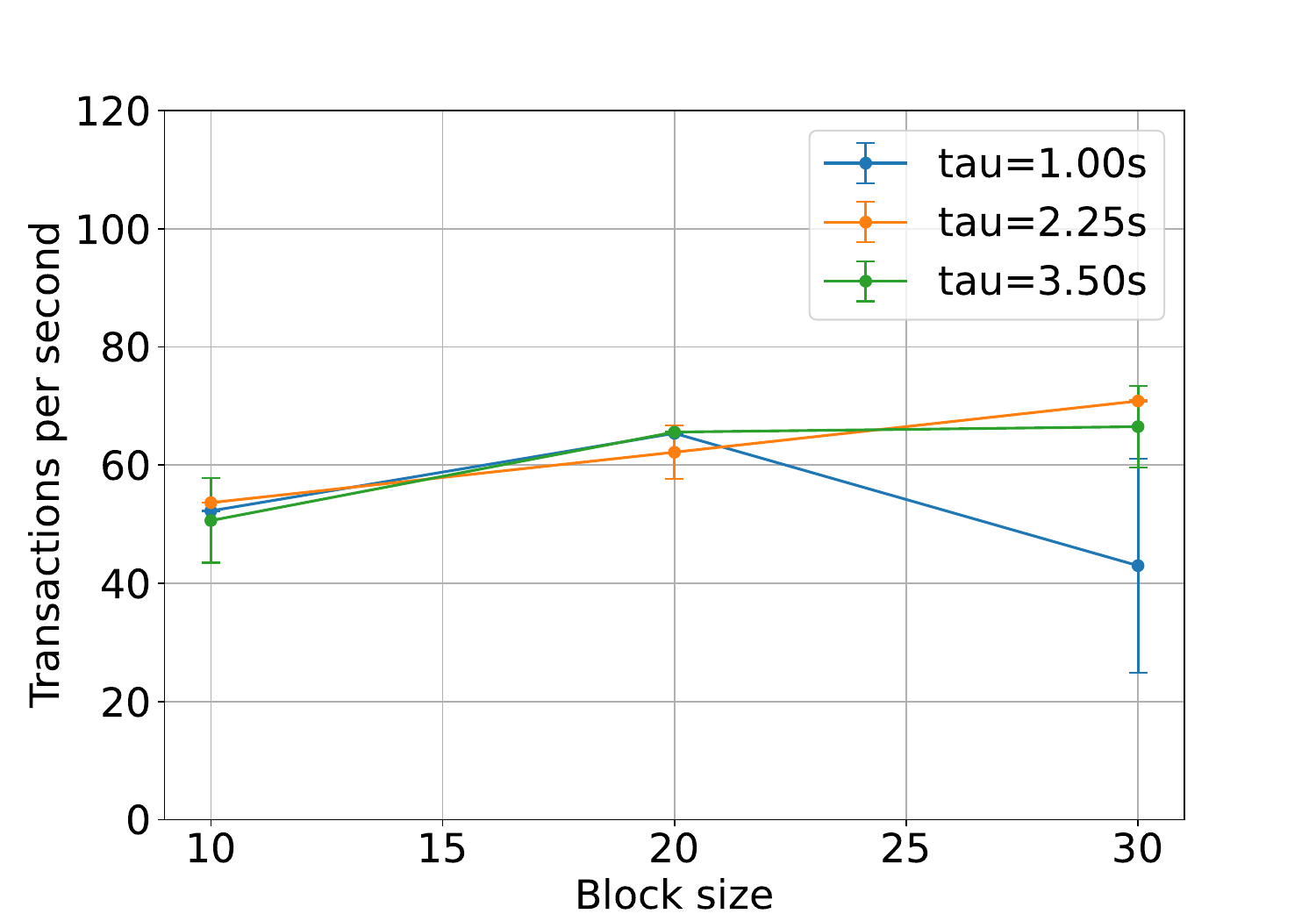}
		\caption{9/9 nodes - TOR anonymization}
	\end{subfigure}
	\begin{subfigure}{0.32\textwidth}
		\includegraphics[width=\textwidth]{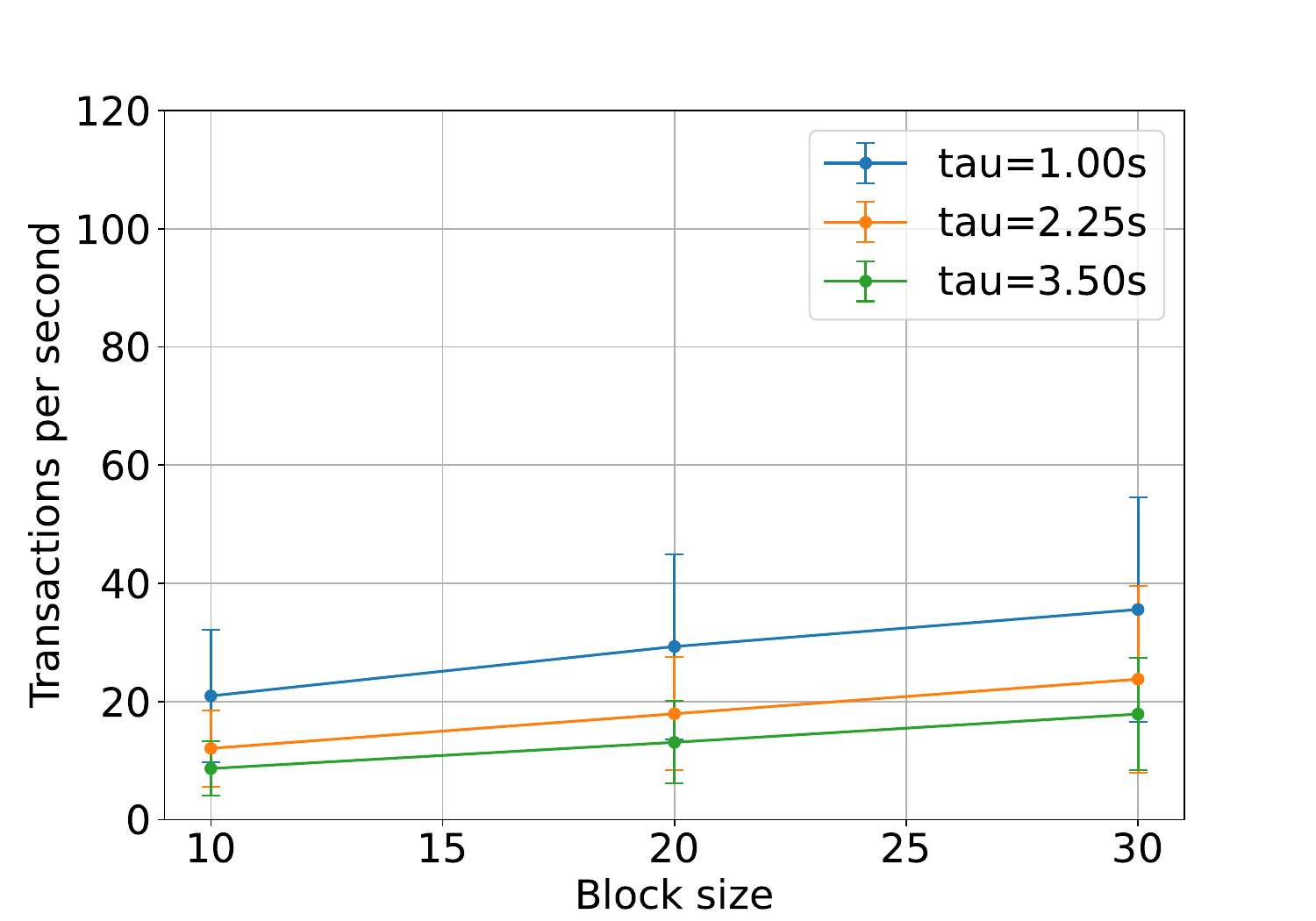}
		\caption{7/9 nodes - TOR anonymization}
	\end{subfigure}
	\caption{Selected experimental results with 9 nodes.}
	\label{fig:exp_nodes_9}
\end{figure*}

\begin{figure*}[!ht]
	\vspace{-0.4cm}
    \centering
	\begin{subfigure}{0.32\textwidth}
		\includegraphics[width=\textwidth]{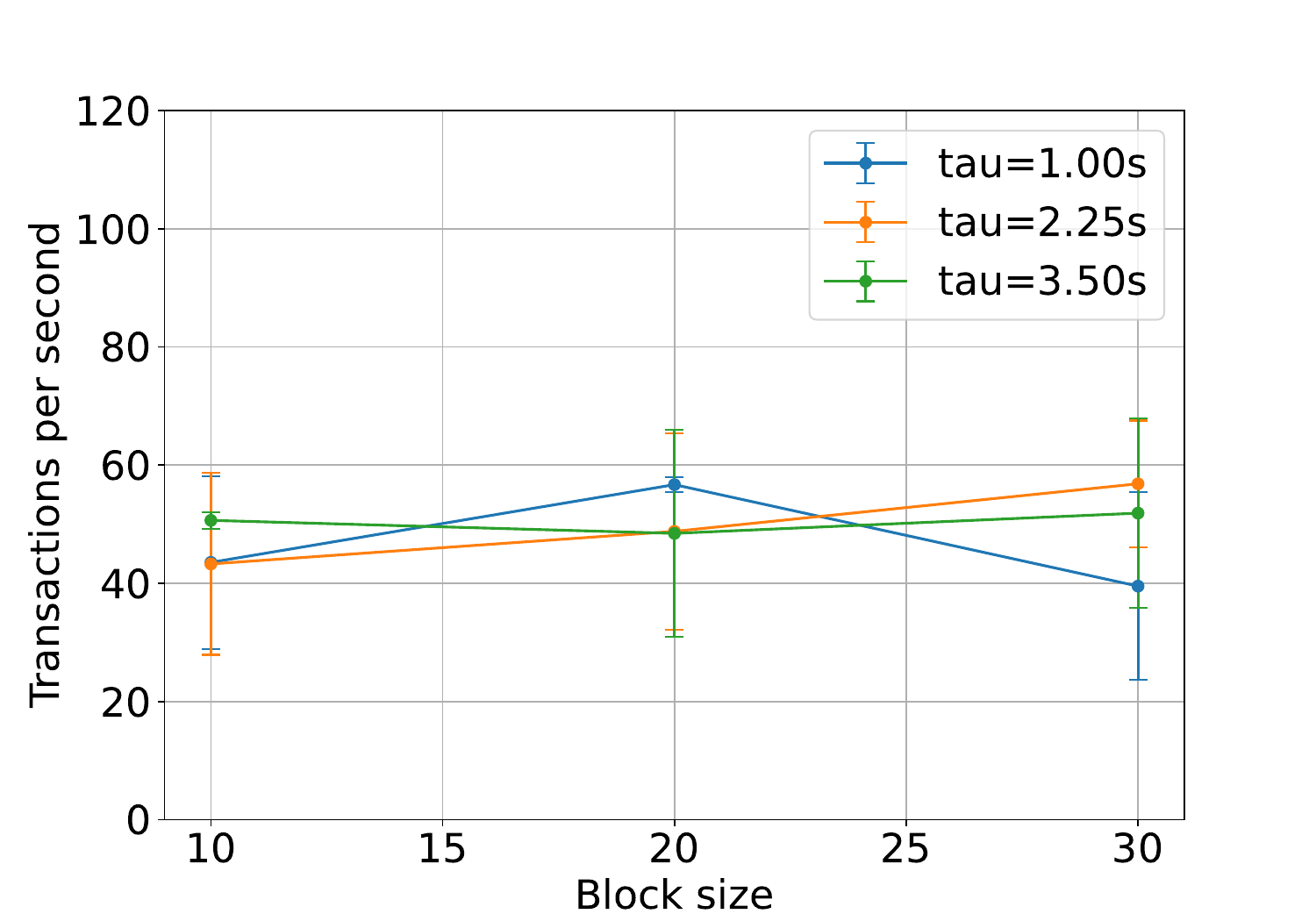}
		\caption{12/12 nodes - no anonymization}
	\end{subfigure}
	\begin{subfigure}{0.32\textwidth}
		\includegraphics[width=\textwidth]{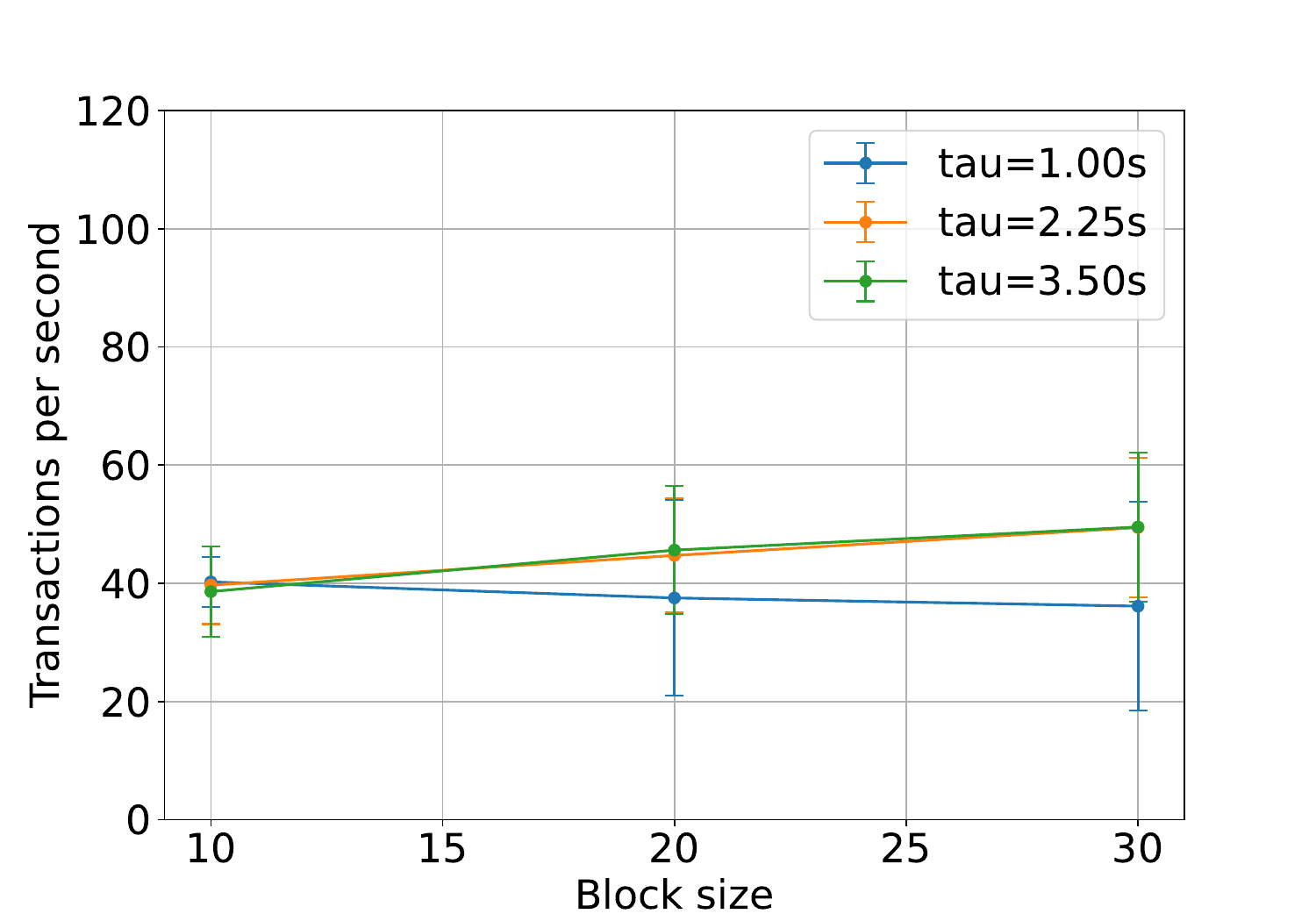}
		\caption{12/12 nodes - TOR anonymization}
	\end{subfigure}
	\begin{subfigure}{0.32\textwidth}
		\includegraphics[width=\textwidth]{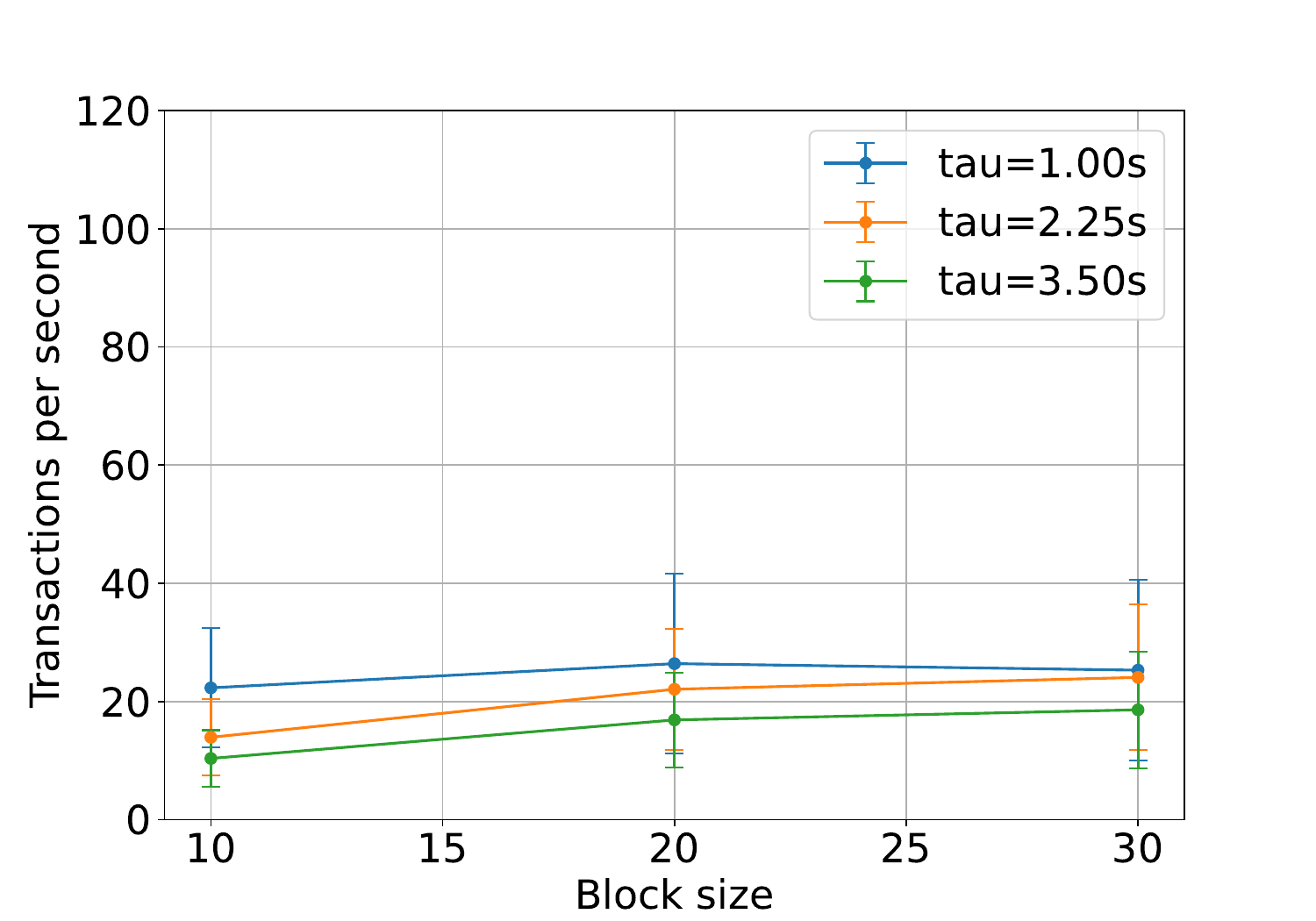}
		\caption{10/12 nodes - TOR anonymization}
	\end{subfigure}
	\caption{Selected experimental results with 12 nodes.}
	\label{fig:exp_nodes_12}
    \vspace{-0.4cm}
\end{figure*}

\section{Evaluation}
\label{sec:eval}
We evaluated the performance of \name through a series of experiments conducted on a local machine with an Intel Core i5 9600KF CPU and 32GB of RAM.
The implementation\footnote{Available at: \url{https://github.com/st22nestrel/COPOR}} was written in Python 3, and the nodes communicated through a localhost socket interface.
We tested the protocol with 6, 9, and 12 nodes, and measured the throughput in transactions per second (tx/s) for each of the three anonymization modes, as well as a baseline without anonymization
We also investigated the impact of node churn by running experiments with one or two nodes offline.

Experiments with 6, 9, and 12 nodes are shown in \autoref{fig:exp_nodes_6}, \autoref{fig:exp_nodes_9}, and \autoref{fig:exp_nodes_12}, respectively.
As shown in \autoref{fig:exp_nodes_6}(a), the baseline performance without anonymization reaches approximately 110 tx/s with a block size of 30 transactions.
\autoref{fig:exp_nodes_6}(b) shows that the introduction of the TOR-like anonymization layer results in a slight performance decrease, as expected.
However, throughput remains relatively high, demonstrating the efficiency of the native anonymization approach. \autoref{fig:exp_nodes_6}(c) illustrates the impact of node churn.
With two nodes offline, the throughput of the TOR-like mode decreases, but the protocol remains operational, highlighting its resilience.
Experiments with 9 and 12 nodes showed similar trends.
As the number of nodes increases, the throughput decreases, but the protocol remains stable.

The Gossip-node and Dandelion modes exhibited slightly lower performance than the TOR-like mode, because of their more complex routing patterns.
In general, our evaluation demonstrates that \name can achieve its design goals.
The native anonymization layer provides effective resistance to DoS with minimal impact on performance, making \name a viable design solution to build high-throughput, scalable and secure PoS blockchains.
\section{Security and Discussion}
\label{sec:security}

\subsection{Liveness and Safety}
CAP theorem~\cite{brewer2000towards} enables a distributed system to select either \textbf{\underline{C}}onsistency or  \textbf{\underline{A}}vailability during the time of network \textbf{\underline{P}}artitions.
If the system selects consistency, it stalls and does not provide liveness (i.e., the blocks are not produced) but provides safety (i.e., all nodes agree on the same blocks when some are produced). 
The example of such a system is Algorand~\cite{algorand} and Streamlet~\cite{chan2020streamlet} due to BFT-like voting.
On the other hand, if the system selects availability, it does not provide safety but provides liveness, which translates into the possibility of creating forks and eventually accepting one as valid.
\name favors liveness over safety alike many PoS consensus protocols~\cite{reijsbergen2020laksa,nakamoto2008bitcoin}.

\medskip \noindent
\subsubsection{\textbf{Safety Violation}}
There is only one situation when safety is not (temporarily) provided by \name, and it happens when the leader goes offline.
If the leader of the round is offline, the new alternative leader producing a block is elected upon expiration of timeout $\tau^B$.
However, the main leader might go on-line before the checkpoint occurs and try to re-build a stronger forked chain with her block included.
Even thought consensus nodes might extend this chain (as they believe it will be ``stronger'' at the epoch's end), there is a chain quality contribution from context sensitive transactions that ``vote'' on the main chain and thus cannot be included in the blocks that do not precede their voted blocks. 
This makes the option of rebuilding a stronger chain less attractive to a rational adversary.

\subsubsection{\textbf{Fairness of Stake}}
Fairness\cite{pass2017fruitchains} is achieved by the fact that a weighed pseudo-random function is used to select the next leader, where the weight is proportional to the peer's stake.

\subsubsection{\textbf{DoS Resistance}}
DoS resistance is achieved by node anonymization -- the ID of the leader in round $r$ is known at the end of round $r-1$ but it is not possible to obtain its network address from the ID. 
Similarly, it is unfeasible for the adversary to DoS all the peering nodes of the victim since the adversary does not know whether they are exit nodes for the victim or just relay nodes on the gossiping path. 

\subsection{Discussion}

\subsubsection{\textbf{Number of Circuits}}
In \name, each consensus node has by default eight peers that it communicates with. 
In TOR, each connection uses 1 circuit. 
Increased number of circuits reduces anonymity, but increases a chance of the successful message delivery. 
After the use of a circuit, the messages are gossiped in the network. This design choice was made to balance the overhead of onion routing with the need for reliable message propagation.

\subsubsection{\textbf{Overhead of Relayed Messages}}
Our protocol imposes a certain network data overhead of the onion-routed messages since every message is encrypted multiple times (by default three times).
Therefore, the initial overhead of the onion-routed message is not negligible. The size of a serialized transaction in our protocol is 192 Bytes, and the size of a serialized block header is 295 Bytes. A message containing a block with 10 transactions is 2229 Bytes.

\subsubsection{\textbf{Deanonymization}}
If the adversary is capable of running a high number of consensus nodes with a minimal stake, there exists a chance that a victim node will select all the nodes of onion routing belonging to the adversary, which could lead to a network layer deanonymization of the victim, and thereby to potential DoS attack on the leader.
Moreover, just one of the (eight) peers is enough to be compromised in this way.

The likelihood of this situation is indirectly proportional to the number of peers $m$ in the circuit and directly proportional to the number of nodes $p$ (i.e., 8 by default) that it has the circuit established with.

\subsubsection{\textbf{Directory}}
It is important to protect against censorship of nodes (concealment of existence of some nodes).
An adversary could use it to conceal the existence of all nodes except those controlled by himself. That would lead to the adversary having control over all the nodes in a circuit, thus de-anonymizing the communication.
The list of peers can be an integral part of blockchain, as in~\cite{directory}, or merely a directory server similar to those in Tor given we can trust it for some reason (e.g. it is controlled by us, its contents is being verified through some mechanism orthogonal to those described in this paper, or we trust the providing party for non-technical reasons, like contractual).

\subsubsection{\textbf{Dynamic Set of Participants}}
For our proof-of-concept implementation, we assumed only static set of participants, to make implementation easier for us. 
However, by implementing neccessary functions to register the new node that wants to join the protocol and functions to obtain the current blockchain state, our approach can be extended to work with dynamic set of participants

\section{Related Work}
\label{sec:related}

\paragraph{\textbf{R3V}}
R3V is a consensus protocol based on round-robin selection mechanism with utilizing Verifiable Delay Function (VDF) \cite{raikwar2021r3v}. The whole consensus is running in Trusted Execution Environment (TEE). To create a block in the network there must exist stakeholders which entitle them to participate in solving VDF. If someone from the eligible stakeholders finds the solution to VDF puzzle then the stakeholder proposes a new block. In this part the round-robin manner comes into the play. Every stakeholder holds a decreasing queue of stakeholders based on their age (the queue is updated every round). The stakeholders will pick a list of predefined size from the queue. If the proposed block is proposed by someone from the list, then the block is accepted and the age of the proposer will update. The round will increase and the solving of the VDF starts again. 

The problem of this algorithm is that an adversary can use DOS attacks on stakeholders and thus can change the acceptance probability of proposed block. Algorithm does not expect  excesive forkness of the chain and the liveness is better with comparison to other state-of-the-art consensus protocols. The throughput cannot be determined because the paper \cite{raikwar2021r3v} does not state any experiment results.

\paragraph{\textbf{Algorand}}
Algorand~\cite{algorand} uses verifiable random functions (VRF) to select new leaders secretly and thus avoids DoS on the leader. VRF is a public-key version of a keyed cryptographic hash. Only the holder of the private key can compute the hash, but anyone with public key can verify the correctness of the hash\cite{vrf}.
Algorand uses VRF for selection of N members of committee by letting the peers compute VRF of a round randomness, selecting those with result lesser than certain value. VRF can’t be used for selection of a single leader. Also, it might happen that none of the peers is selected, causing a synchronization delay to restart the round.

Throughput of Algorand is limitted because each new leader must be confirmed by messages from all members of the committee (similar to BFT protocols), which imposes $N$ additional messages.

Algorand is not a pure BFT protocol, but its hybrid variant that optimizes the number of messages -- one phase of BFT is performed only within a small committee ( < number of participants). If a malicious node is a member of a committee, it might not send a message and if the committee contains a sufficient number of adversarial nodes (i.e., 33\%), the protocol might not produce the block.

\paragraph{\textbf{Tendermint}}
Tendermint is a BFT-based Proof-of-Stake protocol. It has fixed committee members. A block proposer is selected in a round-robin manner~\cite{tendermint}.
This means that a leader is known in advance to all the nodes. An adversary can use this information to perform a DoS attack against the current leader, effectively preventing him from proposing a block. If the adversary does this to every leader other than himself, he can become the only leader to ever propose a block.
Because of BFT, Tendermint has a relatively low throughput. Each round consists of three steps:
\begin{itemize}
	\item Propose - a proposed block is broadcasted
	\item Prevote - peer validates a block and broadcasts its willingness to commit it
	\item Precommit - after receiving information from other peers and if at least $\frac{2}{3}$ peers have validated it, the peer signs the block and commits it in a special commit step
\end{itemize}

The 2nd and 3rd step delay protocol: their full substitute can significantly increase throughput of protocol. Additionally, since the committee is fixed, Tendermint doesn’t scale well.

\paragraph{LaKSA}
LaKSA is derived from Algorand, and adopted ideas from DFINITY and Randhound~\cite{hanke2018dfinity, syta2017scalable}.
It is a proper Proof-of-Stake protocol with some BFT ideas ~\cite{reijsbergen2020laksa}.
It was developed to reduce drawbacks as high reward variance and long confirmation times.
It enhances Algorand properties such as lightweight committee voting. It should be more robust and easily scalable than other protocols.
In LaKSA, committee members are randomly and periodically sampled to vote for their preferred main chain views~\cite{reijsbergen2020laksa}.

The LaKSA can be divided into two rounds: (1) voting for a virtual block, and (2) publishing a new block.

The voters are selected according to a beacon in the last block, similarly to Algorand. The leader that publishes the block is selected by the beacon as well. This two-round schema is not very useful for high throughput.
The authors discuss two possible approaches to protect protocol against DoS attacks on the leaders. They show possible integration of VRF function from Algorand, which enables hiding the node's role, but this approach slows down the block commitment process. The second approach proposes adopting some lightweight mechanism that provides network anonymity. Without additional details, they suggest Dandelion~\cite{bojja2017dandelion} (resp. Dandelion++) as one of the appropriate anonymity protocols. However, this is not embedded into the consensus protocol itself, and thus might not provide sufficient throughput.

\paragraph{Streamlet}
Streamlet~\cite{chan2020streamlet} draws from Casper FFG~\cite{butterin-casper-ffg} and HotStuff~\cite{yin2019hotstuff} but unifies them into the simplest known BFT blockchain paradigm.
It operates in partially synchronous networks and achieves BFT with up to $f < \frac{n}{3}$ faulty nodes among n total nodes, providing deterministic finality without proof-of-work. 
The protocol organizes execution into fixed-length epochs, with a pre-determined leader per epoch selected via a public function (e.g., round-robin). 
In Streamlet, the epoch leader proposes a new block extending the longest notarized chain it knows. 
The block includes the proposal and a justification (votes from prior epochs).
Replicas vote via signed messages only on the first valid proposal received in the epoch, extending their current chain. 
Votes are simple -- a replica votes for a block if it extends the longest notarized prefix and is justified.

\section{Conclusion}
\label{sec:conclusion}

In this paper, we have presented \name, a novel PoS consensus protocol that provides DoS resistance through a native onion routing layer.
By integrating anonymization directly into the consensus protocol, \name is able to protect its leaders from targeted attacks without sacrificing the performance benefits of a single-leader design.
Our experimental evaluation demonstrates that \name can achieve a moderate throughput with a manageable overhead from the anonymization layer, making it a promising approach for building secure, scalable, and resilient PoS blockchains.
The principles and techniques presented in this work offer a path forward for designing next-generation consensus protocols that are both fast and secure.
While the presented results are promising, we acknowledge that the performance was evaluated in a controlled environment.
Future work should focus on a more extensive evaluation on a larger, geographically distributed network to further validate the scalability and real-world performance of \name and additional optimizations.
Additionally, a formal analysis of the probability of deanonymization would provide a more rigorous understanding of the security guarantees of the protocol.

\section*{Acknowledgments}
This work was supported by Brno University of Technology (FIT-S-23-8151), NextGenerationEU (Slovakia project 09I05-03-V02-00057), and the Chips JU DistriMuSe project (Grant 101139769).
Resources were provided by the e-INFRA CZ (ID:90254).

\IEEEtriggeratref{18}
\bibliography{IEEEabrv,references}
\bibliographystyle{IEEEtran}

\end{document}